\documentclass{article}

      \usepackage[preprint]{neurips_2020}

\usepackage[utf8]{inputenc} 
\usepackage[T1]{fontenc}    
\usepackage{hyperref}       
\usepackage{url}            
\usepackage{booktabs}       
\usepackage{amsfonts}       
\usepackage{nicefrac}       
\usepackage{microtype}      
\usepackage{graphicx}       

\title{A Multi-agent Reinforcement Learning Study of Emergence of Social Classes out of Arbitrary Governance: The Role of Environment}

\author{%
  Aslan S.~Dizaji\thanks{The code of this project is available in \url{https://github.com/aslansd/modified-ai-economist}.} \\
  AutocurriculaLab\\
  Tehran, Iran\\
  \texttt{asataryd@umich.edu} \\
}

\begin{document}

\maketitle

\begin{abstract}
  There are several theories in economics regarding the roots or causes of prosperity in a society. One of these theories or hypotheses \textendash named geography hypothesis \textendash mentions that the reason why some countries are prosperous and some others are poor is the geographical location of the countries in the world as makes their climate and environment favorable or unfavorable regarding natural resources. Another competing hypothesis states that man-made institutions particularly inclusive political institutions are the reasons why some countries are prosperous and some others are poor. On the other hand, there is a specific political theory developed for the long-term social development in Iran \textendash named Arbitrary Rule and Aridisolatic Society which particularly emphasizes on the role of aridity to shape arbitrary political and economical institutions in Iran, without any functional social classes in the society. In this paper, by extending the AI-Economist \textendash a recently developed two-level multi-agent reinforcement learning environment \textendash I show that when the central planner is ruling the environment by arbitrary rules, the society evolves through different paths in different environments. In the environment having band-like vertical isolated patches of natural resources, all mobile agents are equally exploited by the central planner and the central planner is also not gaining any income, while in the society having more uniformly distributed natural resources, the productivity and Maximin are higher and the society generates a heterogeneous stratified social structure. All these findings provide a partial answer to the above debate and reconcile the role of geography and political institutions on the long-term development in a region.
\end{abstract}

\section{Introduction}

There are at least three general theories regarding the roots of prosperity in a society. One theory named geography hypothesis mentions that the reason why some countries are prosperous and some others are poor is due to the location of the countries in the world which makes them particularly amenable to use and extract natural resources or makes them unsuited. This second hypothesis called culture hypothesis mentions that the reason why some countries are rich and some others are poor is due to the inherent features of the people living in those countries considering their culture, religion, or ethnicity which make them particularly responsive to the ideas of work ethic, progress, and innovation or not. Finally, the third theory mentions that the reason behind prosperity of some countries and poverty in others is due to the fact that the leaders of poor countries are ignorant about how to rule their countries to guide their nations toward prosperity. All these theories are backed with some historical data, but all of them simultaneously can be refuted by counter-examples \citep{Acemoglu2012}.

More recently there is a new class of general theories regarding the nature and cause of prosperity in a country which points towards the importance of inclusive economic and particularly political institutions in generating a prosperous or less prosperous future for a country. This theory which is backed by Daron Acemoglu and colleagues mentions that inclusive institutions which make the economic and political landscape flat for all groups in the society \textendash so motivate them to participate in fair economic and political activities \textendash make a nation or a country prosperous \citep{Acemoglu2011, Acemoglu2012, Acemoglu2015a, Acemoglu2020, Acemoglu2022a, Acemoglu2022b}.

On the other hand, there is a specific theory for the long-term social development in Iran named Arbitrary Rule and Aridisolatic Society developed by Homa Katouzian \citep{Katouzian2003} which indicates that the nature of social institutions in Iran is completely different from their counterparts in the west. As an instance, using historical data, Katouzian indicates that in Iran the state is the only functional entity of the society which is completely independent from all urban social classes, while there is not any functional urban social class which its identity is independent from the state, and as a result, all urban classes are more or less empirical. Moreover, while there is a large body of laws, but due to lack of functional urban classes, there is not any non-violable binding rules between the state and the society which makes the government essentially arbitrary. Katouzian attributes the emergence of the arbitrary governance to the aridity in the vast region of Iranian plateau. The aridity generates large number of small isolated villages which their individual surplus is not sufficient to found a feudal base but the sum of whole surplus of these villages could make an arbitrary governance with large transportation facilities across the country and infrastructure in the urban areas. Then the arbitrary rule could make all urban social classes dependent on itself while perpetuates its power across country until that point that some internal or external conditions ignite revolution. At this point, since the arbitrary governance is independent from any social classes, all urban classes are less or more against the government.

Here, I bring the summary of the Katouzian’s theory using his own words: “To sum up, aridity did play a basic role in shaping the structure of the Iranian political economy and its institutional features, but it did so (to borrow Tolstoy’s words) in its own peculiar way: (a) it served to create autonomous village units of production, none of which could produce a sufficiently large surplus to provide a feudal power base and (b) but, given the expanses of the region, the collective surplus of all these isolated and autonomous villages taken together was so large that, once taken by an external force, it could be used as the economic source of a countrywide despotic power. The despotic apparatus could then impose itself and its arbitrary will on all the social classes, and prevent the subsequent fragmentation of politiconomic power until such time that a combination of internal and/or external pressures would destroy it, and \textendash sooner or later \textendash replace it by another despotic apparatus. The size of the direct and indirect collective agricultural surplus was so large as to enable these despotic states to spend on transport, communications, military and bureaucratic organizations, and so on, which both maintained their hold on the land and prevented the later emergence of feudal autonomy in agriculture, or bourgeois citizenship in towns.” \citep{Katouzian2003}

In this paper, I intend to reconcile the two theories posed by Homa Katouzian and Daron Acemoglu by showing the interplay of natural environment and the resultant institutions. I perform this by extending the AI-Economist framework \citep{Zheng2022}, a recently developed two-level multi-agent reinforcement learning environment. In this framework, one single agent is a rational social planner who designs a particular mechanism or policy generally having a goal of optimizing a particular kind of social welfare functions in the society. The other agents are a set of rational economic agents who behave in response to the implemented mechanism or policy generally following their own self-interest. This framework has been used to model the tax-gaming behavior of agents \textendash optimizing their labors, trading, and building, while the central social planner maximizes productivity or equality in the society \citep{Zheng2022}. I explained and used the extension of the AI-Economist, the Modified AI-Economist, in an accompanying paper \citep{Dizaji2023a} in which I show the impacts of the governing systems or institutions on the origin of morality, prosperity or equality, and fairness in the society. Here using the same framework, but considering two parallel environments in which one of them is comprised of band-like isolated and the other one of uniformly distributed natural resources, I intend to show that if the central planner is an arbitrary ruler, each environment evolves through a different path. Band-like environment finally converges to an environment in which all the agents are getting powerless in front of the naked power of the arbitrary governance, while the central planner's net total tax revenue is also getting zero. On the other hand, the uniform environment converges to a final situation in which the society is getting composed of stratified distinct social classes, and the central planner is also able to continue collecting the non-zero taxes. In the band-like environment, while the final Equality is higher (basically all the agents are equally exploited by the central government), the Productivity and Maximin are lower than the uniform environment. This interesting result obtained considering the fact that the total amounts of natural resources in the band-like environment are slightly more than the total amounts of natural resources in the uniform environment. Furthermore, the arbitrary nature of the central planner is devised by letting it to return some arbitrary amounts of tax values to more wealthier agents. Overall, this paper is another manifestation of the power of multi-agent reinforcement learning to model social and economical phenomena \citep{Trott2021, Zheng2022, Zhang2022, Leibo2019, Leibo2021, Johanson2022}.

\section{Modified AI-Economist}
For a complete description of the AI-Economist, please refer to the original paper (\citep{Zheng2022}) and Appendix A. Here, three major modifications that are made to the original framework for the purpose of this paper are explained. 

\begin{figure}
	\centering
	\includegraphics[width=0.7\linewidth]{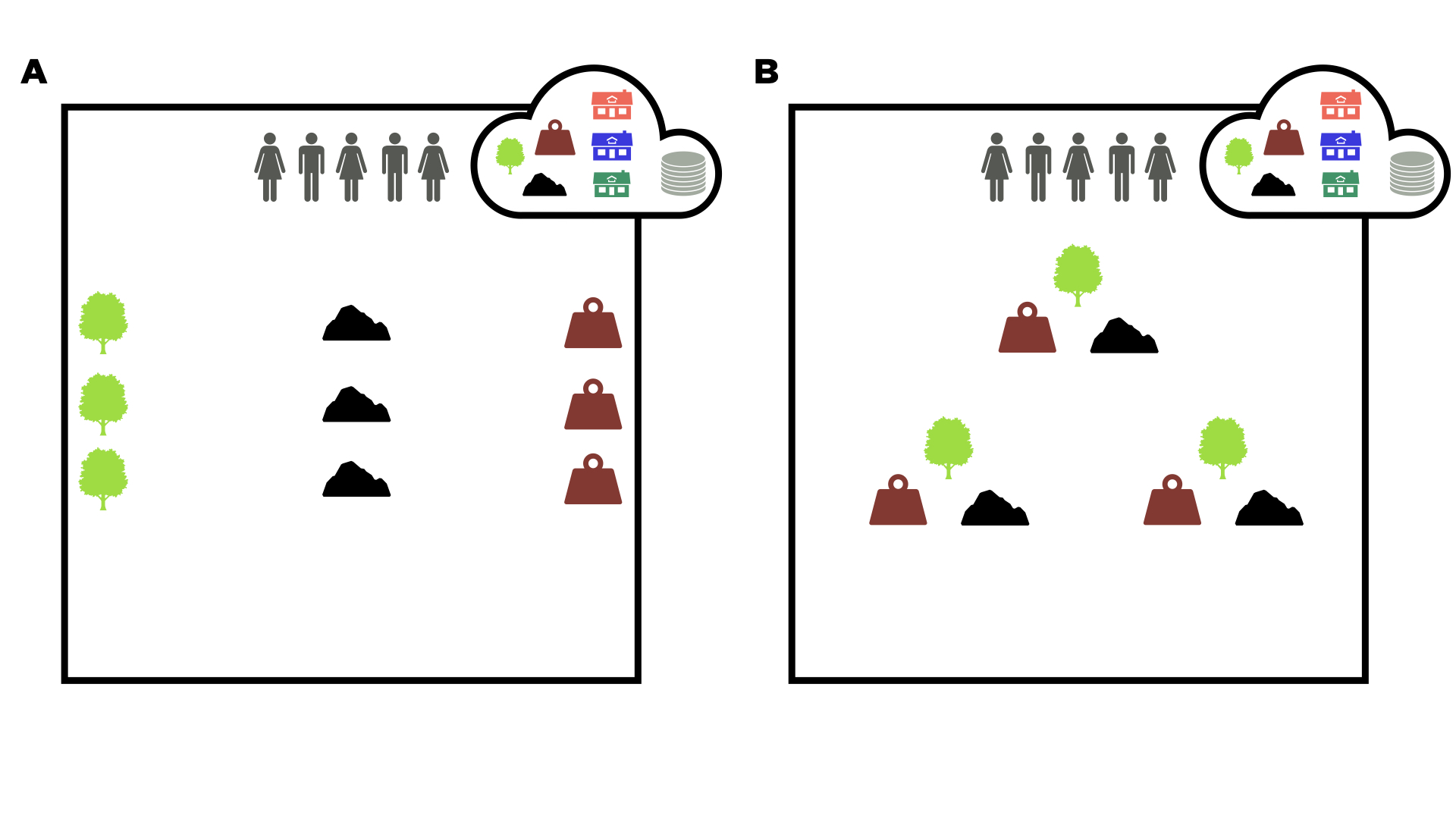}
	\caption{A schematic figure showing the two environments of the Modified AI-Economist used in this paper. In all simulations of this paper, there are 5 agents in the environment which simultaneously cooperate and compete to gather and trade three natural resources, using them to build houses and earn incomes, and at the end of each tax period, pay their taxes to the central planner. The central planner optimizes its own reward function which could be a combination of equality and productivity in the society, and returns arbitrary partial amounts of the total collected taxes to the mobile agents.}
	\label{Figure1}
\end{figure}

First, one new resource material \textendash iron \textendash is added to the environment, and now with three building materials, the number of possible house types is diversified to three: a red house is made of wood and stone, a blue house is made of wood and iron, and a green house is made of stone and iron. Also, three different build-skills for three different house types are introduced. However, move and gather skill and labor, and trade labor will be equal and fixed across all materials and agents. Also, in the modified version of the AI-Economist, the social planner is able to observe the complete public map of the environment (Fig.~\ref{Figure1}).

\begin{figure}
	\centering
	\includegraphics[width=0.7\linewidth]{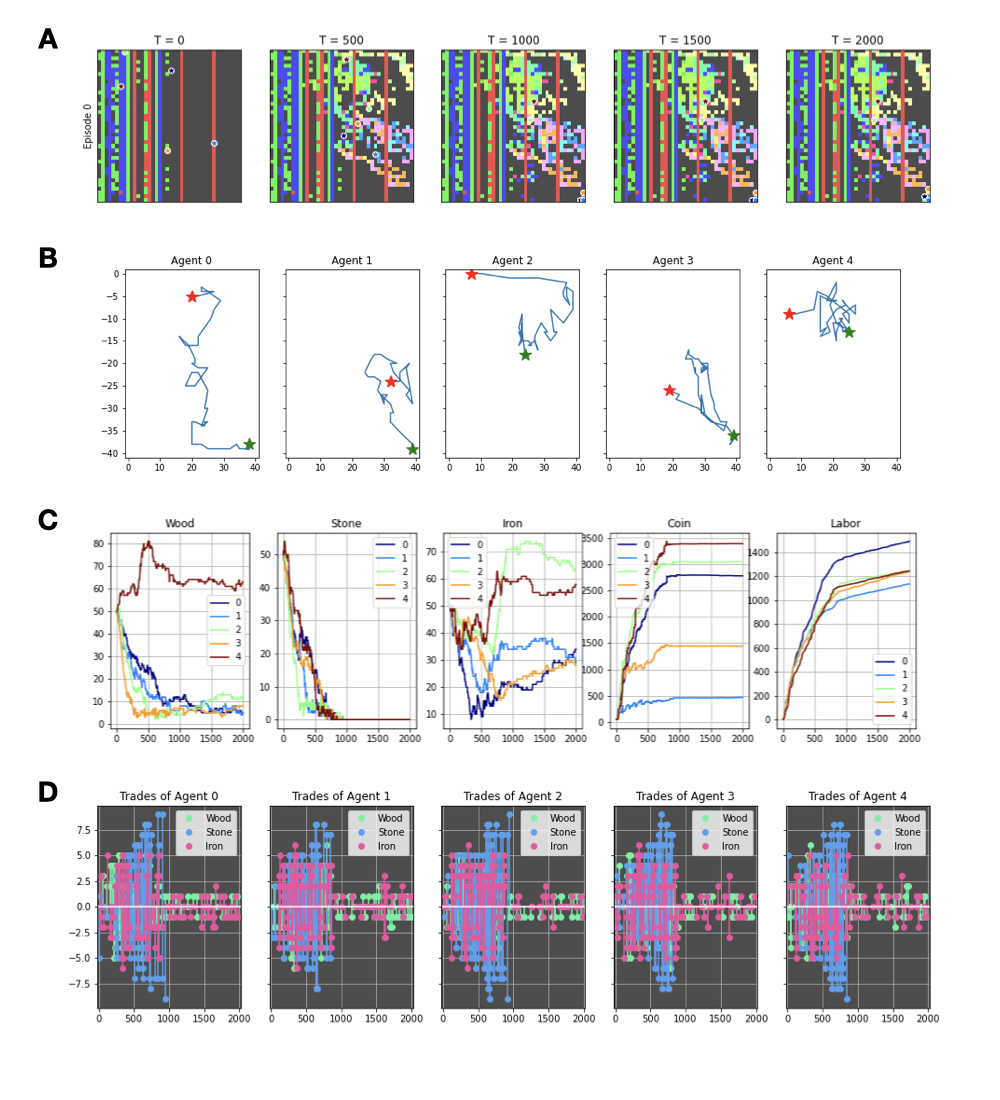}
	\caption{Sample plots obtained from running the Modified AI-Economist in the band-like environment with equality times productivity as the objective function of the central planner. (A) The environment across five time-points of an episode, (B) the movement of the agents across an episode, (C) the budgets of three resources plus coin and labor of the agents across an episode, (D) and the trades of three resources of the agents across an episode. As it is clear, at some point during an episode, the agents cannot earn more incomes (panel (C), the fourth plot from the left) and the trades of resources also largely decrease (panel (D)).}
	\label{Figure2}
\end{figure}

\begin{figure}
	\centering
    \includegraphics[width=0.7\linewidth]{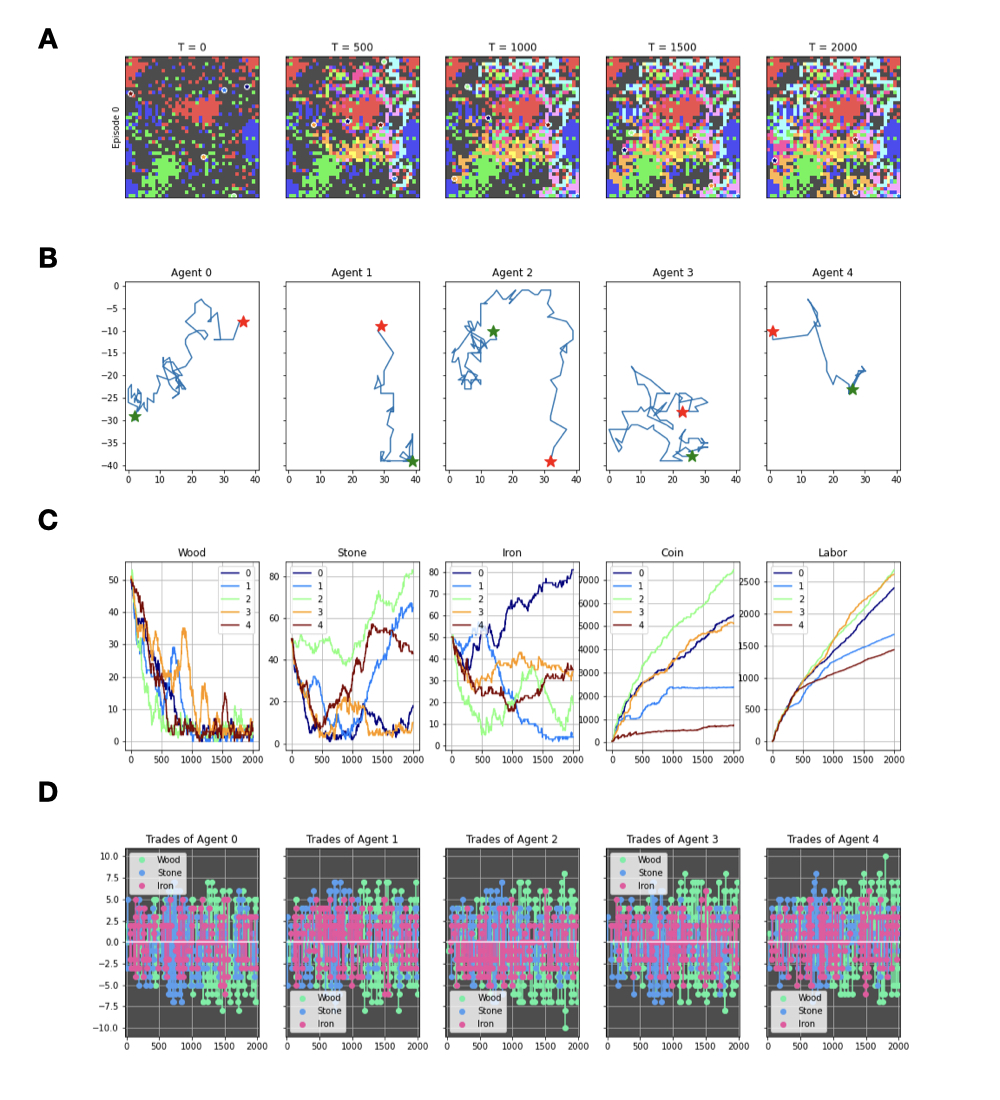}
    \caption{Sample plots obtained from running the Modified AI-Economist in the uniform environment with equality times productivity as the objective function of the central planner. (A) The environment across five time-points of an episode, (B) the movement of the agents across an episode, (C) the budgets of three resources plus coin and labor of the agents across an episode, (D) and the trades of three resources of the agents across an episode. As it is clear, the agents are able to earn more incomes (panel (C), the fourth plot from the left) and the trades of resources are also unchanged across an episode (panel (D)).}
    \label{Figure3}
\end{figure}

Second, for the purpose of this paper, there are two different environments. The first one is composed of band-like vertical isolated patches of unique natural resources placed in the whole environment (Fig.~\ref{Figure2}(A)), and the second one is composed of uniformly distributed natural resources placed in the environment (Fig.~\ref{Figure3}(A)).

Third, the central planner collects the taxes as it is mentioned in the original framework of the AI-Economist, but here it returns arbitrary partial amounts of the net total tax revenue to more wealthier agents of the environment using the following two formula. In the following, \( urn \) refers to a uniform random number, \( nti \) refers to the net total income of all agents, \( noa \) refers to the number of all agents, \( nowa \) refers to the number of wealthier agents. \( ri \) refers to a random integer, and finally \( nttr \) refers to the net total tax revenue of the central planner.

\begin{equation}\label{Equation1}
	Last\:Income[Agent] > (0.7 + 0.1 * urn) * \frac{nti}{noa}
\end{equation}

\begin{equation}\label{Equation2}
    Return\:Tax[Agent] = \frac{noa}{nowa} * (1 + (-1)^{ri} * urn * (1 - \frac{nowa}{noa})) * \frac{nttr}{noa}
\end{equation}

In the original framework of the AI-Economist, the net total tax revenue of the central planner is equally divided across all the agents while in this modified framework, the net total tax revenue of the social planner is somewhat arbitrarily divided among a group of pre-selected wealthier agents \textendash having incomes in the previous tax period more than partially random limits (Eq.~\ref{Equation1}). Thus the central planner here is arbitrary and none-inclusive, and its discriminative power is in favor of the wealthier agents in the society. Simultaneously, Eq.~\ref{Equation2} is devised by assuming that as an episode progresses, slowly, all the agents are getting incomes more than the pre-specified random limits, and thus they are included in the tax return scheme of the central planner. In that case, all the agents get an equal share of the net total tax revenue. Overall, these formula \textendash in an ideal situation \textendash can model the emergence of equality before the law (\citep{Acemoglu2020}). In the next section, it is indicated that this assumption is partially confirmed though by showing an interesting distinct patterns of convergence in two environments.

\section{Results}

\begin{figure}
	\centering
	\includegraphics[width=0.7\linewidth]{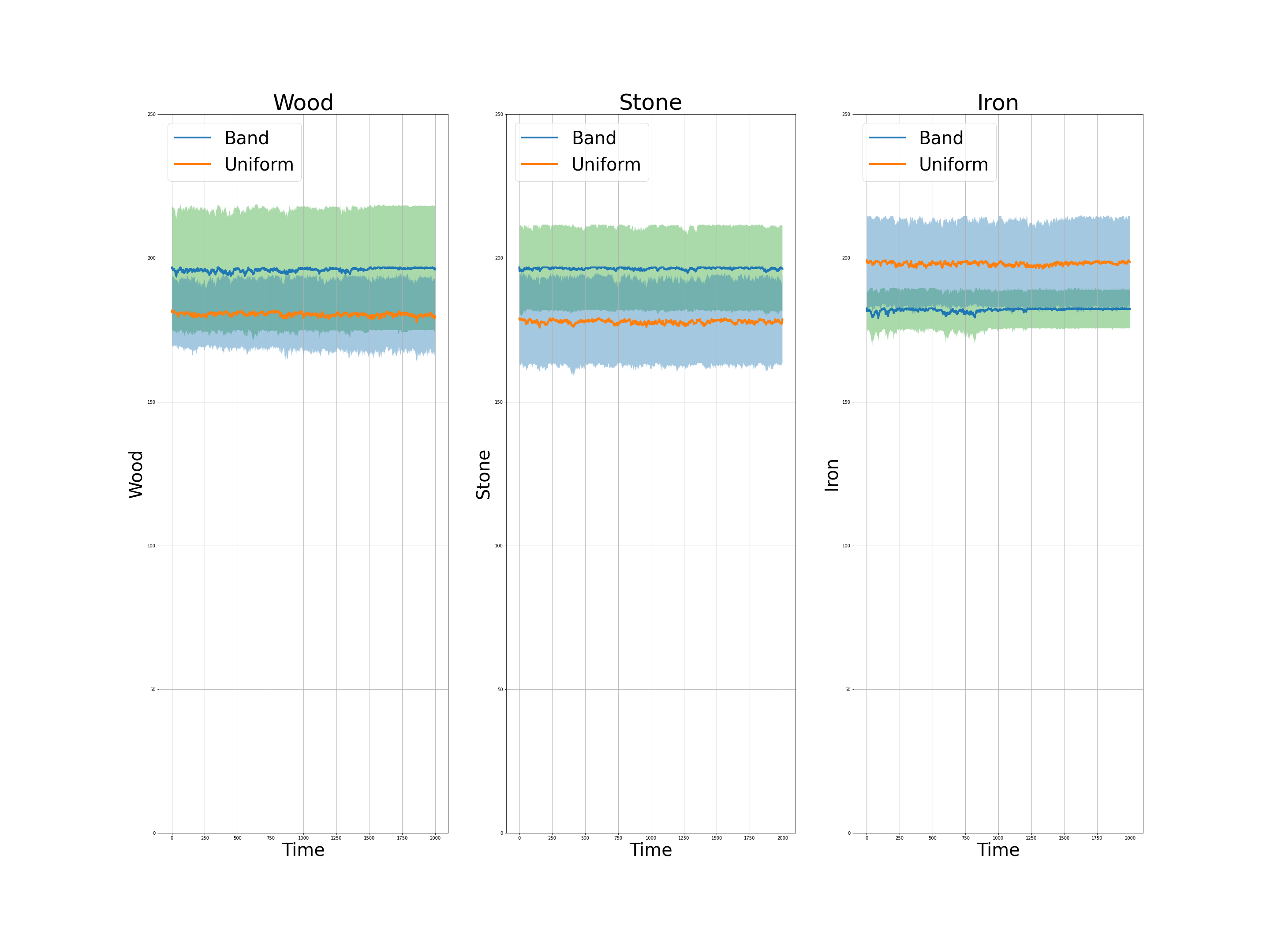}
	\caption{The level of natural resources across the two environments, band-like and uniform. As it is clear from the plots, this level remains almost constant across one episode and overall it is slightly higher for band-like environment for two out of three natural resources compared to the uniform environment.}
	\label{Figure4}
\end{figure}

\begin{figure}
	\centering
	\includegraphics[width=0.7\linewidth]{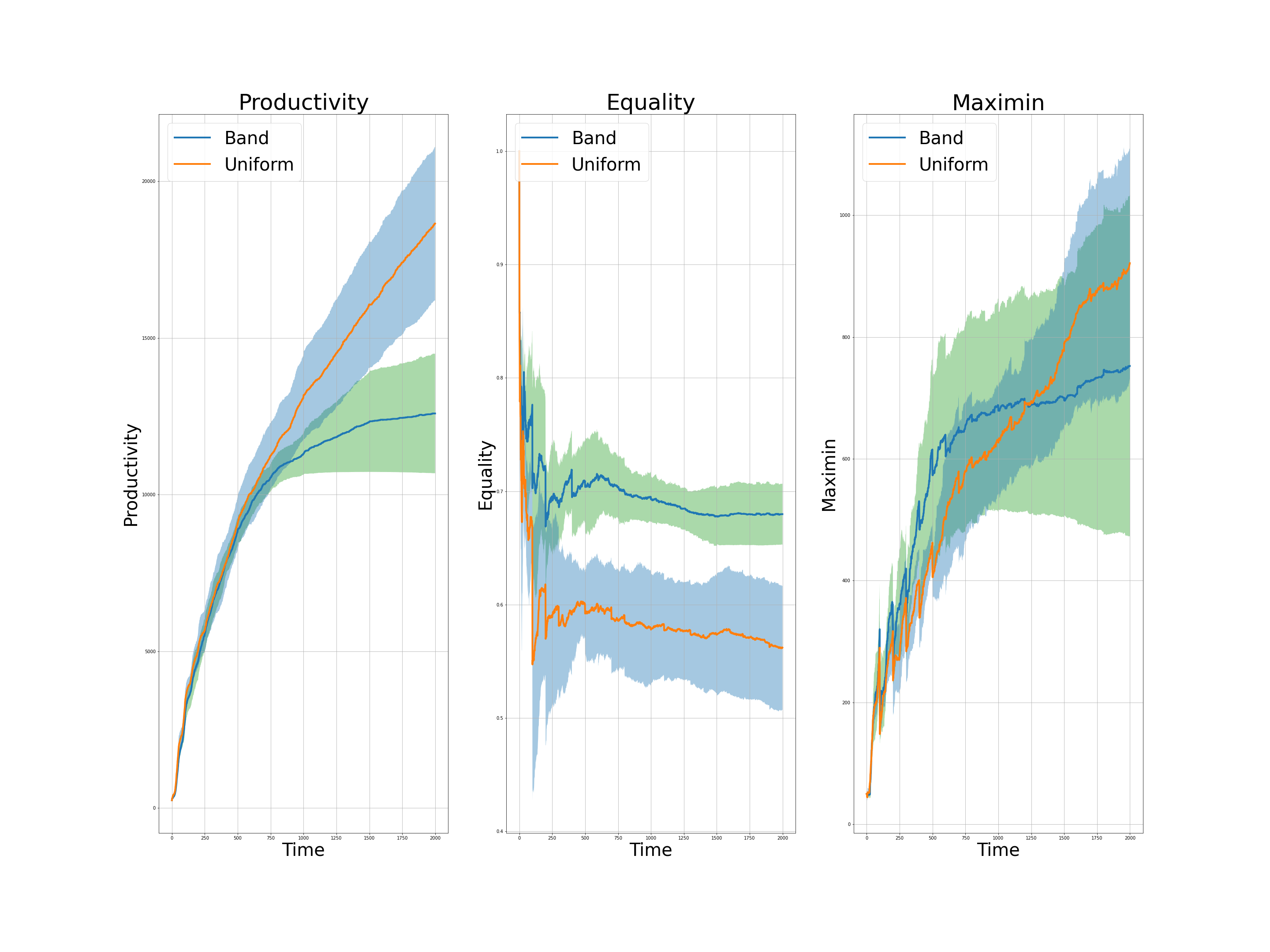}
	\caption{Productivity, Equality, and Maximin across the two environments, band-like and uniform. As it is clear from the plots, the Productivity and Maximin are higher while the Equality is lower in the uniform compared to the band-like environment.}
	\label{Figure5}
\end{figure}

\begin{figure}
	\centering
	\includegraphics[width=0.7\linewidth]{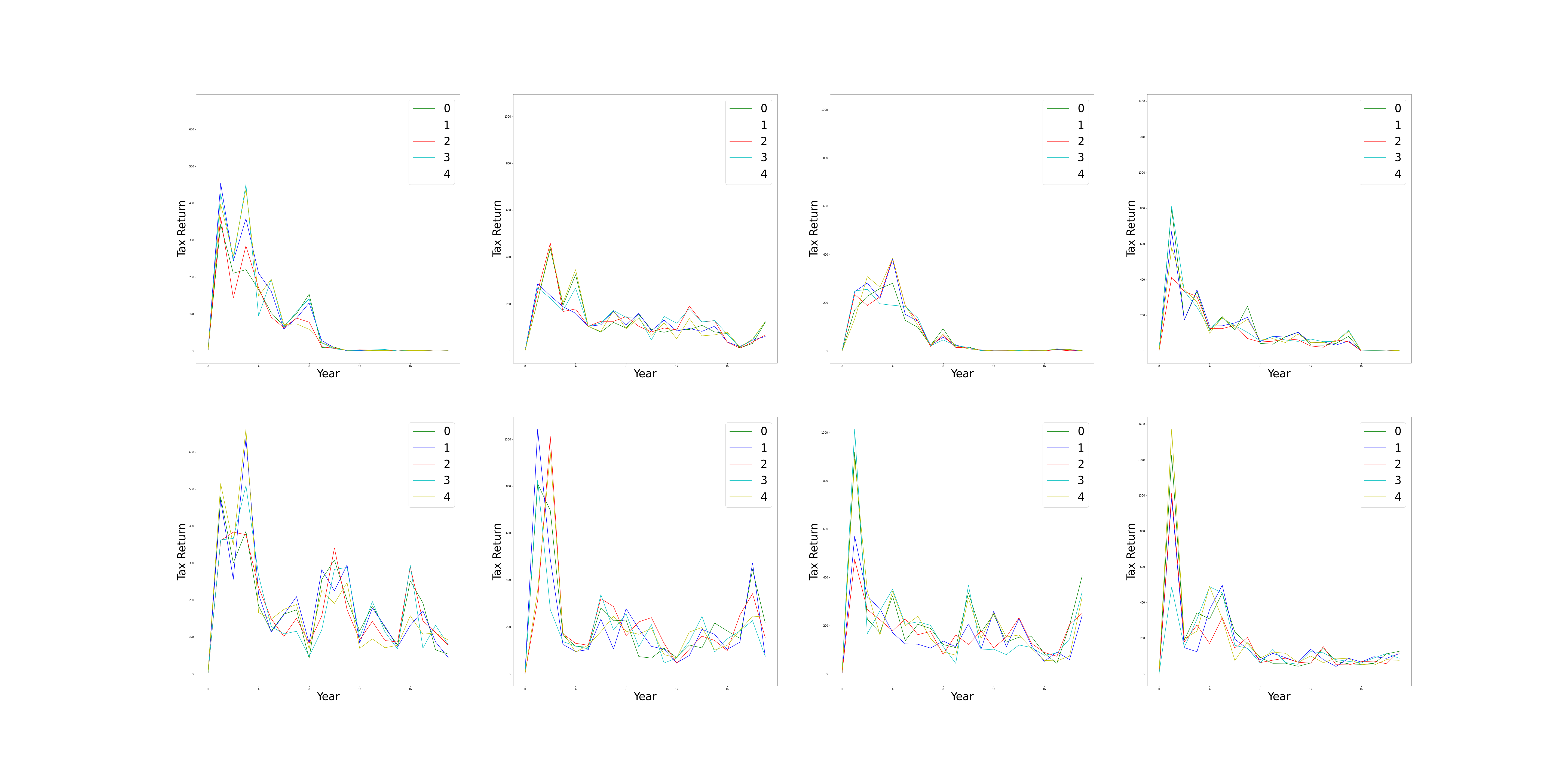}
	\caption{Amount of tax return to all agents across 20 years of one simulation for all 8 runs of this paper with the order of Fig.~\ref{Figure9} from left-to-right and top-to-bottom. The top-row shows the plots for 4 simulations of the band-like environment, while the bottom-row shows the plots for 4 simulations of the uniform environment. As it is clear from the plots, in three out of four plots of the band-like environment, at some point during an episode, the amounts of tax returns to all agents are getting zero due to the reason that the net total income of all agents and thus the central planner's net total tax revenue are both getting zero. This situation does not happen for the agents in the uniform environment, and in all plots until the end of an episode, they keep their socially stratified structure.}
	\label{Figure6}
\end{figure}

Fig.~\ref{Figure4} shows the amounts of three natural resources in the environment across an episode. It is clear from this plot, that the level of natural resources in the two environments remains constant while this level is slightly higher in the band-like environment compared to the uniform environment for two out of three natural resources. Fig.~\ref{Figure5} shows that the Productivity and Maximin are higher in the uniform compared to the band-like environment, while the Equality is lower. Finally, Fig.~\ref{Figure6} shows the amounts of tax return by the central planner to all agents across an episode for all 8 runs of this paper depicted in Fig.~\ref{Figure9}. The top-row shows the plots for the band-like environment in which the amounts of tax return are getting zero for all agents as an episode progresses for three out of four simulations. Moreover, the bottom-row shows the plots for the uniform environment in which the amounts of tax return are not getting zero across an episode for all four simulations. All these plots (from Fig.~\ref{Figure2} to Fig.~\ref{Figure6}) speak to the following facts: in the band-like environment with isolated vertical patches of unique natural resources, as an episode continues the agents are earning less and less incomes until that point that their net total income is getting zero. Simultaneously, the net total tax revenue of the central planner is also getting decreased and equal to zero. Basically, in this environment at the end of one episode, all the agents are equally exploited by the central planner and there is not any difference among them. The central planner is also not gaining anything further from the agents, and we could say that the system is failed. On the other hand, in the uniform environment, the net total income of all agents is not getting zero, and as a result, the central planner's net total tax revenue is also not getting zero across an episode. Basically, until the end of an episode, the agents keep their stratified social structure while the central planner is also able to continue collecting the taxes. This is the reason why in the uniform environment, the Equality is lower than the band-like environment, while both Productivity and Maximin are higher. The whole point of this paper is that aridity as it is manifested by the band-like environment is more prone to generate exploitative central planner and a society without any functional social classes, while more favorable environment such as the uniform environment is inclined to generate less exploitative central planner with more stratified social classes. Overall, the model used in this paper is a small step toward reconciling the interplay of environments and institutions on the long-term development of a region.          

\section{Final Remarks}

\paragraph{Current Limitations} There are at least three limitations to the current study. The first limitation comes from the fact that for each set of input parameters of the Modified AI-Economist, only one simulation has been run to generate one set of results. Then the similar runs are pooled together to have average results across different conditions. Thus it is reasonable to run multiple times a simulation with a unique set of parameters and then average them all together. The second limitation is the number of episodes which each training has been run and is equal to 5000. While even with this amount of episodes, the average reward plots across training iterations (Fig.~\ref{Figure10}) show that almost all simulations have been converged, it is wise to try more RL iterations. Finally, the third limitation comes from the fact that the discriminative nature of the central planner in this paper is only due to the arbitrary tax return scheme, however, the central planner has an objective function treating all mobile agents equally which has been considered unchanged across the two environments. One immediate modification to the current modeling, is to include partially the agents in the objective function of the central planner to see how this changes the dynamics in the two environments.

\paragraph{Future Directions} Beside above modifications, two other important extensions for this project can be envisioned. The first one is to test other values for the constants in Eq.~\ref{Equation1} or other modeling frameworks for Eqs.~\ref{Equation1} and ~\ref{Equation2}. This way we could make sure that the results obtained here are robust. The second direction is to model punishment in this framework. One important feature of arbitrary governance is that there is not any non-violable set of laws between the central planner and the mobile agents even considering the case of punishment, so it would be interesting to model this phenomenon in the general framework of the AI-Economist.

\section*{Broader Impact}
As I discussed in the accompanying paper \citep{Dizaji2023a}, for the works similar to the current paper, it is possible to envision many policy implications, however, we should be cautions about interpreting these results more than what is appropriate, due to many simplifying assumptions inherent in any mathematical modeling. Overall, the limited modeling framework used in this paper shows that in the discussion of the economics development we should consider the roles of geography and political institutions together, and also compare the results of Productivity, Equality, and Maximin cautiously across different environments. 

For further information, please refer to the Ethics section of the original AI-Economist paper \citep{Zheng2022}. Particularly, in that section, it has been emphasized on the full-transparency of the codes of a project with a similar scope. As a result, I provided an open Github repository (\url{https://github.com/aslansd/modified-ai-economist}) having all the required codes, simulations, and notebooks to generate the runs and plots of this paper.

\begin{ack}
\textit{AutocurriculaLab} has been funded in March 2022 and since then has been supported by multiple agencies. Hereby, I acknowledge their supports.
\end{ack}

\medskip
\small

\bibliography{Multi-agent_Reinforcement_Learning_for_Economics_and_Neuroscience}
\bibliographystyle{apalike}

\newpage

\section{Appendix A: AI-Economist}
Here, a detailed description of the original AI-Economist is brought \citep{Zheng2022}:

\begin{enumerate}
	\item The AI-Economist is a two-level deep RL framework for policy design in which agents and a social planner co-adapt. In particular, the AI-Economist uses structured curriculum learning to stabilize the challenging two-level, co-adaptive learning problem. This framework has been validated in the domain of taxation. In two-level spatiotemporal economies, the AI-Economist has substantially improved both utilitarian social welfare and the trade-off between equality and productivity over baselines. It was successful to do this despite emergent tax-gaming strategies, accounting the emergent labor specialization, agent interactions, and behavioral changes.
	
	\item Stabilizing the training process in two-level RL is difficult. To overcome, the training procedure in the AI-Economist has two important features \textendash curriculum learning and entropy-based regularization. Both of them encourage the agents and the social planner to co-adopt gradually and not stopping exploration too early during training and getting trapped in local minima. Furthermore, the AI-Economist is a game of imperfect (the agents and the social planner do not have access to the perfect state of the world) but complete (the agents and the social planner know the exact rules of the game) information.
	
	\item The Gather-Trade-Build economy of the AI-Economist is a two-dimensional spatiotemporal economy with agents who move, gather resources (stone and wood), trade them, and build houses. Each agent has a varied house build-skill which sets how much income an agent receives from building a house. Build-skill is distributed according to a Pareto distribution. As a result, the utility-maximizing agents learn to specialize their behaviors based on their build-skill. Agents with low build-skill become gatherers: they earn income by gathering and selling resources. Agents with high build-skill become builders: they learn that it is more profitable to buy resources and then build houses.
	
	\item The Open-Quadrant environment of the Gather-Trade-Build economy has four regions delineated by impassable water with passageways connecting each quadrant. Quadrants contain different combinations of resources: both stone and wood, only stone, only wood, or nothing. Agents can freely access all quadrants, if not blocked by objects or other agents. This scenario uses a fixed set of build-skill based on a clipped Pareto distribution and determine each agent’s starting location based on its assigned build-skill. The Open-Quadrant scenario assigns agents to a particular corner of the map, with similarly skilled agents being placed in the same starting quadrant (Agents in the lowest build-skill quartile start in the wood quadrant; those in the second quartile start in the stone quadrant; those in the third quartile start in the quadrant with both resources; and agents in the highest build-skill quartile start in the empty quadrant).
	
	\item The state of the world is represented as an \( n_{h} \times n_{w} \times n_{c} \) tensor, where \( n_{h} \) and \( n_{w} \) are the size of the world and \( n_{c} \) is the number of unique entities that may occupy a cell, and the value of a given element indicates which entity is occupying the associated location. The action space of the agents includes four movement actions: up, down, left, and right. Agents are restricted from moving onto cells that are occupied by another agent, a water tile, or another agent’s house. Stone and wood stochastically spawn on special resource regeneration cells. Agents can gather these resources by moving to populated resource cells. After harvesting, resource cells remain empty until new resources spawn. By default, agents collect one resource unit, with the possibility of a bonus unit also being collected, the probability of which is determined by the agent’s gather-skill. Resources and coins are accounted for in each agent’s endowment \( x \), which represents how many coins, stone, and wood each agent owns.
	
	\item Agent's observations include the state of their own endowment (wood, stone, and coin), their own build-skill level, and a view of the world state tensor within an egocentric spatial window. The experiment use a world of 25 by 25 for 4-agent and 40 by 40 for 10-agent environments, where agent spatial observations have size 11 by 11 and are padded as needed when the observation window extends beyond the world grid. The planner observations include each agent’s endowment but not build-skill level. The planner does not observe the spatial state of the world.
	
	\item Agents can buy and sell resources from one another through a continuous double-auction. Agents can submit asks (the number of coins they are willing to accept) or bids (how much they are willing to pay) in exchange for one unit of wood or stone. The action space of the agents includes 44 actions for trading, representing the combination of 11 price levels (0, ..., 10 coins), 2 directions (bids and asks), and 2 resources (wood and stone). Each trade action maps to a single order (i.e., bid three coins for one wood, ask for five coins in exchange for one stone, etc.). Once an order is submitted, it remains open until either it is matched (in which case a trade occurs) or it expires (after 50 time steps). Agents are restricted from having more than five open orders for each resource and are restricted from placing orders that they cannot complete (they cannot bid with more coins than they have and cannot submit asks for resources that they do not have). A bid/ask pair forms a valid trade if they are for the same resource and the bid price matches or exceeds the ask price. When a new order is received, it is compared against complementary orders to identify potential valid trades. When a single bid (ask) could be paired with multiple existing asks (bids), priority is given to the ask (bid) with the lowest (highest) price; in the event of ties, priority then is given to the earliest order and then at random. Once a match is identified, the trade is executed using the price of whichever order was placed first. For example, if the market receives a new bid that offers eight coins for one stone and the market has two open asks offering one stone for three coins and one stone for seven coins, received in that order, the market would pair the bid with the first ask and a trade would be executed for one stone at a price of three coins. The bidder loses three coins and gains one stone; the asker loses one stone and gains three coins. Once a bid and ask are paired and the trade is executed, both orders are removed. The state of the market is captured by the number of outstanding bids and asks at each price level for each resource. Agents observe these counts for both their own bids/asks and the cumulative bids/asks of other agents. The planner observes the cumulative bids/asks of all agents. In addition, both the agents and the planner observe historical information from the market: the average trading price for each resource, as well as the number of trades at each price level.
	
	\item Agents can choose to spend one unit of wood and one unit of stone to build a house, and this places a house tile at the agent’s current location and earns the agent some number of coins. Agents are restricted from building on source cells as well as locations where a house already exists. The number of coins earned per house is identical to an agent’s build-skill, a numeric value between 10 and 30. Hence, agents can earn between 10 and 30 coins per house built. Build-skill is heterogeneous across agents and does not change during an episode. Each agent’s action space includes one action for building. Over the course of an episode of 1000 time steps, agents accumulate labor cost, which reflects the amount of effort associated with their actions. Each type of action (moving, gathering, trading, and building) is associated with a specific labor cost. All agents experience the same labor costs.
	
	\item Simulations are run in episodes of 1000 time steps, which is subdivided into 10 tax periods or tax years, each lasting 100 time steps. Taxation is implemented using income brackets and bracket tax rates. All taxation is anonymous: Tax rates and brackets do not depend on the identity of taxpayers. On the first time step of each tax year, the marginal tax rates are set that will be used to collect taxes when the tax year ends. For taxes controlled by the deep neural network of the social planner, the action space of the planner is divided into 7 action subspaces, one for each tax bracket: \( (0, 0.05, 0.10, ..., 1.0)^{7} \). Each subspace denotes the set of discretized marginal tax rates available to the planner. Discretization of tax rates only applies to deep learning networks, enabling standard techniques for RL with discrete actions. The income bracket cutoffs are fixed. Each agent observes the current tax rates, indicators of the temporal progress of the current tax year, and the set of sorted and anonymized incomes the agents reported in the previous tax year. In addition to this global tax information, each agent also observes the marginal rate at the level of income it has earned within the current tax year so far. The planner also observes this global tax information, as well as the non-anonymized incomes and marginal tax rate (at these incomes) of each agent in the previous tax year.
	
	\item The payable tax for income \( z \) is computed as follows:
	\begin{equation}\label{Equation3}
		T(z) = \sum_{j = 1}^{B} \tau_{j} \cdot ((b_{j + 1} - b_{j}) \mathbf{1} [z > b_{j + 1}] + (z - b_{j}) \mathbf{1} [b_{j} < z \leq b_{j + 1}])
	\end{equation}
	where \( B \) is the number of brackets, and \( \tau_{j} \) and \( b_{j} \) are marginal tax rates and income boundaries of the brackets, respectively.
	
	\item An agent’s pretax income \( z_{i} \) for a given tax year is defined simply as the change in its coin endowment \( C_{i} \) since the start of the year. Accordingly, taxes are collected at the end of each tax year by subtracting \( T(z_{i}) \) from \( C_{i} \). Taxes are used to redistribute wealth: the total tax revenue is evenly redistributed back to the agents. In total, at the end of each tax year, the coin endowment for agent \( i \) changes according to \( \bigtriangleup C_{i} = - T(z_{i}) + \frac{1}{N} \sum_{j = 1}^{N} T(z_{j}) \), where \( N \) is the number of agents. Through this mechanism, agents may gain coin when they receive more through redistribution than they pay in taxes. Following optimal taxation theory, agent utilities depend positively on accumulated coin \( C_{i, t} \), which only depends on post-tax income \( \tilde{z}  = z - T(z) \). In contrast, the utility for agent \( i \) depends negatively on accumulated labor \( L_{i, t} = \sum_{k = 0}^{t} l_{i, k} \) at time step \( t \). The utility for an agent \( i \) is:
	\begin{equation}\label{Equation4}
		u_{i, t} = \frac{C^{1 - \eta}_{i, t} - 1}{1 - \eta} - L_{i, t}, \eta > 0
	\end{equation}

	\item Agents learn behaviors that maximize their expected total discounted utility for an episode. It is found that build-skill is a substantial determinant of behavior; agents’ gather-skill empirically does not affect optimal behaviors in our settings. All of the experiments use a fixed set of build-skills, which, along with labor costs, are roughly calibrated so that (i) agents need to be strategic in how they choose to earn income and (ii) the shape of the resulting income distribution roughly matches that of the 2018 U.S. economy with trained optimal agent behaviors.
	
	\item RL provides a flexible way to simultaneously optimize and model the behavioral effects of tax policies. RL is instantiated at two levels, that is, for two types of actors: training agent behavioral policy models and a taxation policy model for the social planner. Each actor’s behavioral policy is trained using deep RL, which learns the weights \( \theta_{i} \) of a neural network \( \pi(a_{i, t} | o_{i, t}; \theta_{i}) \) that maps an actor’s observations to actions. Network weights are trained to maximize the expected total discounted reward of the output actions. Specifically, for an agent \( i \) using a behavioral policy \( \pi_{i}(a_{t} | o_{t}; \theta_{i}) \), the RL training objective is (omitting the tax policy \( \pi_{p} \)):
	\begin{equation}\label{Equation5}
		\max_{\pi_{i}} E_{a_{1} \sim \pi_{1}, ..., a_{N} \sim \pi_{N}, s^{'} \sim P} [\sum^{H}_{t = 0} \gamma^{t} r_{t}]
	\end{equation}
	where \( s^{'} \) is the next state and \( P \) denotes the dynamics of the environment. The objective for the planner policy \( \pi_{p} \) is similar. Standard model-free policy gradient methods update the policy weights \( \theta_{i} \) using (variations of):
	\begin{equation}\label{Equation6}
		\mathbf{\triangle \theta_{i}} \propto E_{{a_{1} \sim \pi_{1}, ..., a_{N} \sim \pi_{N}, s^{'} \sim P}}[\sum^{H}_{t = 0} \gamma^{t} r_{t} \nabla_{\theta_{i}} \log \pi_{i}(a_{i, t} | o_{i, t}; \mathbf{\theta_{i}})]
	\end{equation}

	\item In this work, the proximal policy gradients (PPO) is used to train all actors (both agents and planner). To improve learning efficiency, a single-agent policy network \( \pi(a_{i, t} | o_{i, t}; \theta) \) is trained whose weights are shared by all agents, that is, \( \theta_{i} = \theta \). This network is still able to embed diverse, agent-specific behaviors by conditioning on agent-specific observations.
	
	\item At each time step \( t \), each agent observes the following: its nearby spatial surroundings; its current endowment (stone, wood, and coin); private characteristics, such as its building skill; the state of the markets for trading resources; and a description of the current tax rates. These observations form the inputs to the policy network, which uses a combination of convolutional, fully connected, and recurrent layers to represent spatial, non-spatial, and historical information, respectively. For recurrent components, each agent maintains its own hidden state. The policy network for the social planner follows a similar construction but differs somewhat in the information it observes. Specifically, at each time step, the planner policy observes the following: the current inventories of each agent, the state of the resource markets, and a description of the current tax rates. The planner cannot directly observe private information such as an agent’s skill level.
	
	\item Rational economic agents train their policy \( \pi_{i} \) to optimize their total discounted utility over time while experiencing tax rates \( \tau \) set by the planner’s policy \( \pi_{p} \). The agent training objective is:
	\begin{equation}\label{Equation7}
		\forall i : \max_{\pi_{i}} E_{\tau \sim \pi_{p}, a_{i} \sim \pi_{i}, \mathbf{a_{-i}} \sim \mathbf{\pi_{-i}}, s^{'} \sim P} [\sum^{H}_{t = 1} \gamma^{t} r_{i, t} + u_{i, 0}], r_{i, t} = u_{i, t} - u_{i, t - 1}
	\end{equation}
	where the instantaneous reward \( r_{i, t} \) is the marginal utility for agent \( i \) at time step \( t \). Bold-faced quantities denote vectors, and the subscript \( -i \) denotes quantities for all agents except for \( i \).
	
	\item For an agent population with monetary endowments \( \mathbf{C_{t}} = (C_{1, t}, ..., C_{N, t}) \), the equality \( eq(\mathbf{C_{t}}) \) is defined as:
	\begin{equation}\label{Equation8}
		eq(\mathbf{C}_{t}) = 1 - \frac{N}{N - 1} gini(\mathbf{C}_{t}), 0 \leq eq(\mathbf{C}_{t}) \leq 1
	\end{equation}
	where the Gini index is defined as:
	\begin{equation}\label{Equation9}
		gini(\mathbf{C}_{t}) = \frac{\sum_{i = 1}^{N} \sum_{j = 1}^{N} |C_{i, t} - C_{j, t}|}{2N \sum^{N}_{i = 1} C_{i, t}}, 0 \leq gini(\mathbf{C}_{t}) \leq \frac{N - 1}{N}
	\end{equation}

	\item The productivity is defined as the sum of all incomes:
	\begin{equation}\label{Equation10}
		prod(\mathbf{C}_{t}) = \sum_{i} C_{i, t}
	\end{equation}
	The economy is closed: subsidies are always redistributed evenly among agents, and no tax money leaves the system. Hence, the sum of pretax and post-tax incomes is the same. The planner trains its policy \( \pi_{p} \) to optimize social welfare:
	\begin{equation}\label{Equation11}
		\max_{\pi_{p}} E_{\tau \sim \pi_{p}, \mathbf{a} \sim \mathbf{\pi}, s^{'} \sim P} [\sum^{H}_{t = 1} \gamma^{t} r_{p, t} + swf_{0}], r_{p, t} = swf_{t} - swf_{t - 1}
	\end{equation}

	\item The utilitarian social welfare objective is the family of linear-weighted sums of agent utilities, defined for weights \( \omega_{i} \geq 0 \):
	\begin{equation}\label{Equation12}
		swf_{t} = \sum^{N}_{i = 1} \mathbf{\omega}_{i} \cdot \mathbf{u}_{i, t}
	\end{equation}
	Inverse-income is used as the weights: \( \omega_{i} \propto \frac{1}{C_{i}} \), normalized to sum to one. An objective function is defined that optimizes a trade-off between equality and productivity, defined as the product of equality and productivity:
	\begin{equation}\label{Equation13}
		swf_{t} = eq(\mathbf{C}_{t}) \cdot prod(\mathbf{C}_{t})
	\end{equation}
\end{enumerate}

\newpage

\section{Appendix B: Supplemental Figures}

\begin{figure*}[h!]
	\centering
	\includegraphics[width=0.7\linewidth]{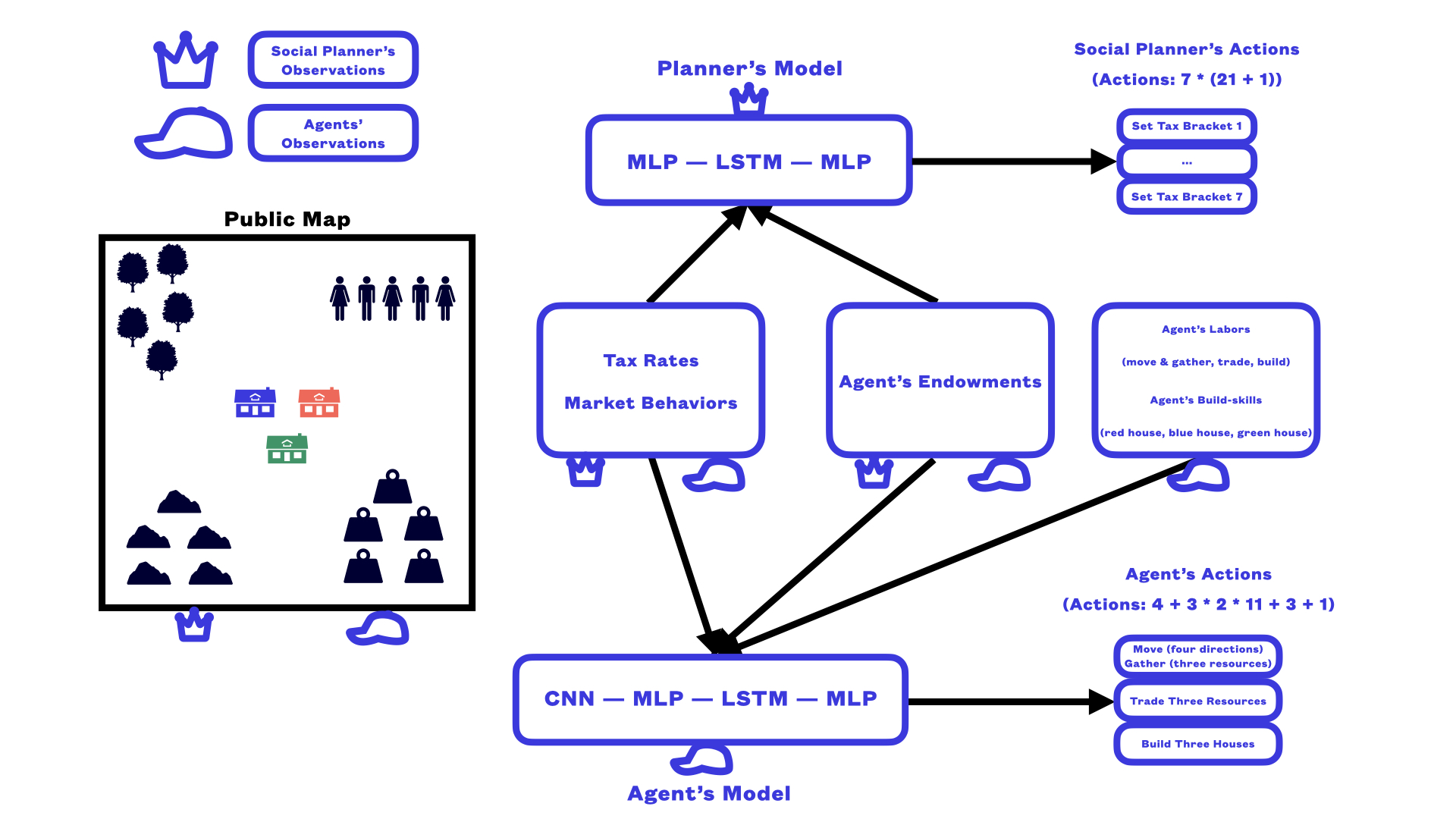}
	\caption{Observation and action spaces for economic agents and the social planner. The agents and the planner observe different subsets of the world state. Agents observe their spatial neighborhood, market prices, tax rates, inventories, labor, and skill level. Agents can decide to move (and therefore gather if moving onto a resource), buy, sell, or build. There are maximum 74 unique actions available to the agents. The planner observes the public spatial map, market prices, tax rates, and agent inventories. The social planner in both environments decides how to set tax rates, choosing one of 22 settings for each of the 7 tax brackets. MLP: multi-layer perceptron, LSTM: long short-term memory, CNN: convolutional neural network. This figure should be compared to Fig. 9 of the original AI-Economist paper \citep{Zheng2022}.}
	\label{Figure7}
\end{figure*}

\newpage

\begin{figure*}[h!]
	\centering
	\includegraphics[width=0.7\linewidth]{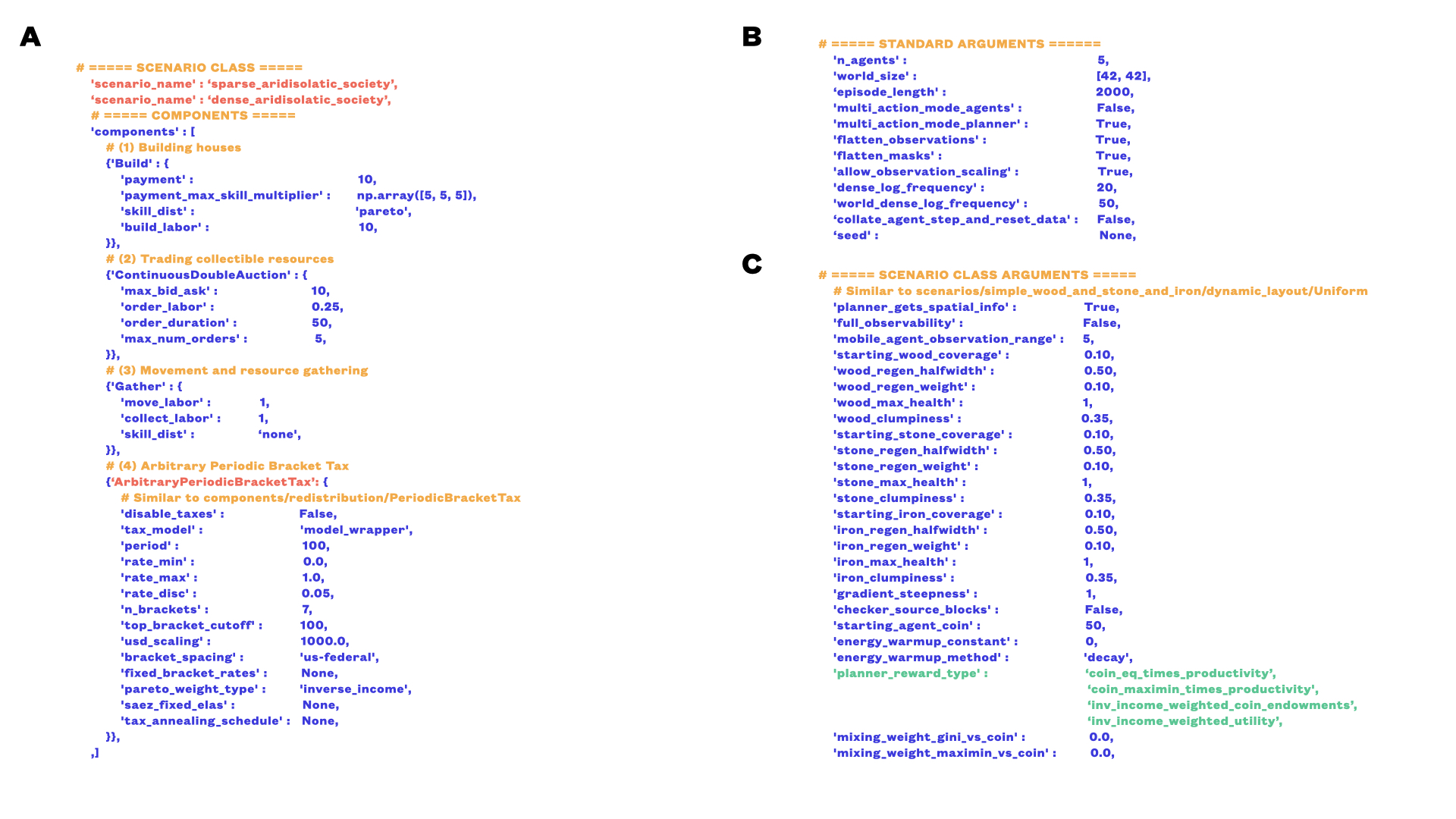}
	\caption{(A)(B)(C) Different features and input parameters of the Modified AI-Economist framework which have been tested and their aggregated plots are brought and discussed in the main text. The orange lines indicate various parts of the input structure. The two red lines in the top left indicate the two possible environments, band-like (sparse) and uniform (dense) tested in this paper, and the red line in the middle left points out the arbitrary tax schedule. On the right, the four green lines indicate four possible reward functions of the central planner.}
	\label{Figure8}
\end{figure*}

\newpage

\begin{figure*}[h!]
	\centering
	\includegraphics[width=0.7\linewidth]{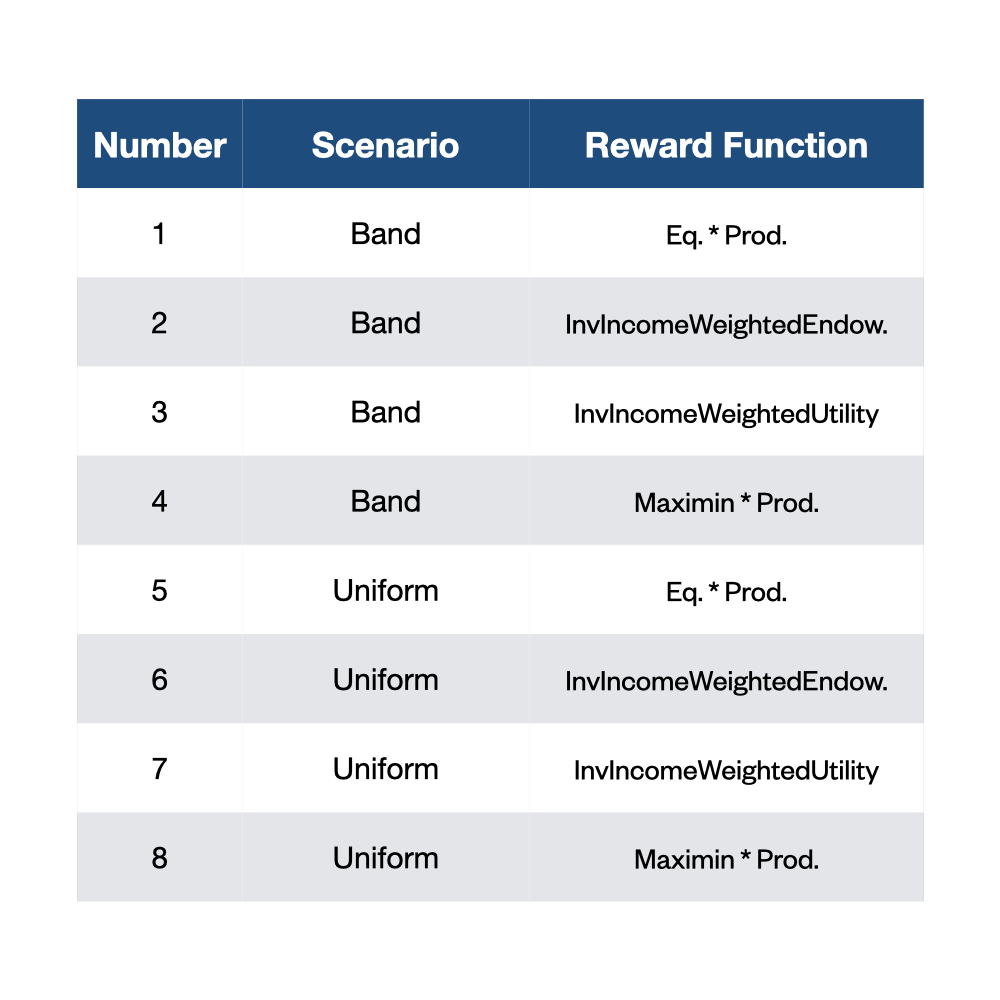}
	\caption{A figure showing all different runs of the Modified AI-Economist environment with different values as input parameters. The \textit{Reward Function} refers to the reward function of the central planner. To generate the plots in the main text, the generated results of all simulations belonging to one environment are pooled together.}
	\label{Figure9}
\end{figure*}

\newpage

\begin{figure}
	\centering
	\includegraphics[width=0.7\linewidth]{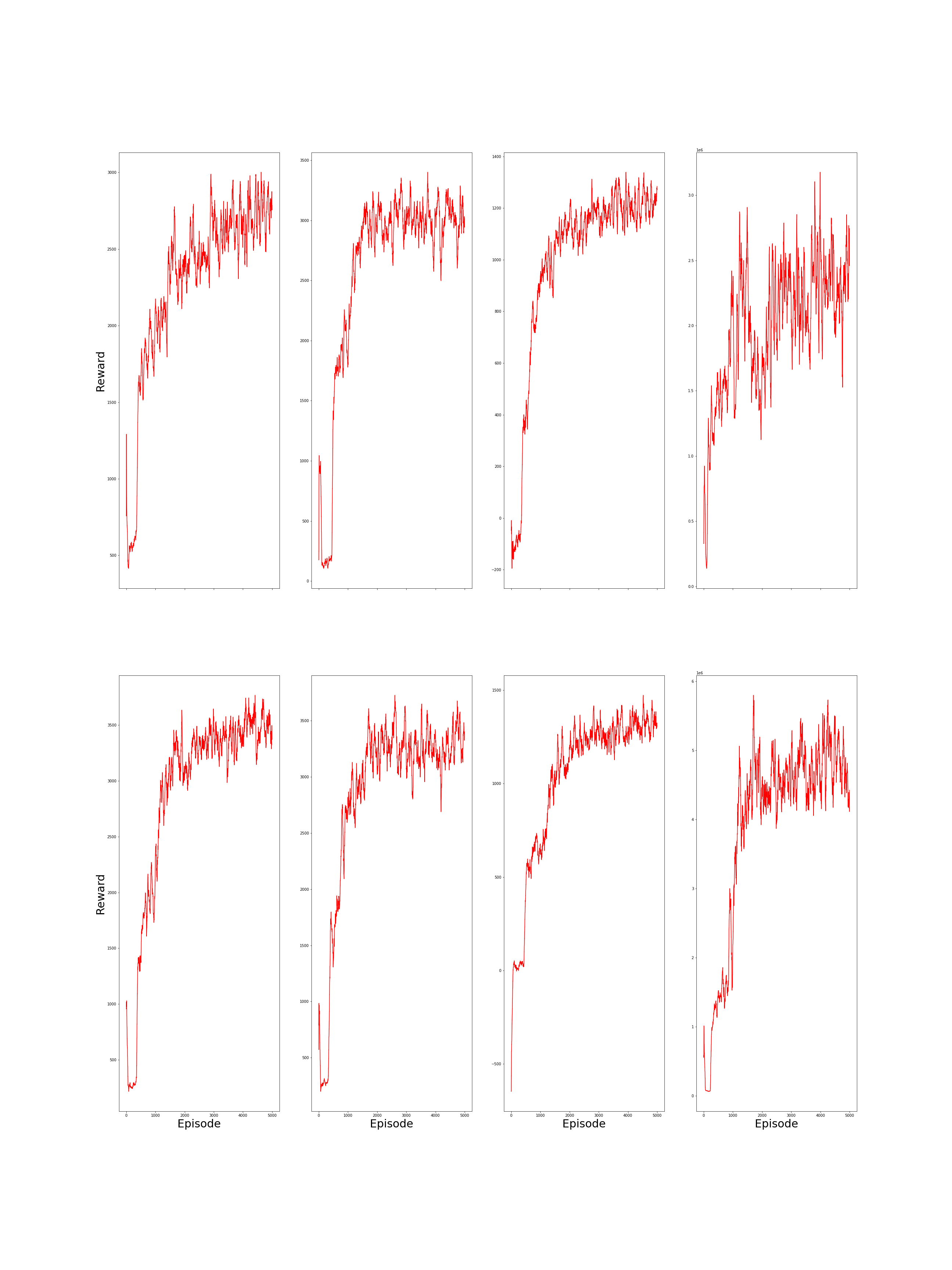}
	\caption{Average episode reward across training \textendash 5000 episodes \textendash for all runs of this paper. The plots of the 8 runs of the Fig.~\ref{Figure9} are brought in the order from left-to-right and top-to-bottom. It is worthwhile to mention that the training of two-level RL is particularly unstable, but it seems that almost all the simulations have been converged.}
	\label{Figure10}
\end{figure}

\end{document}